# *In-situ* measurements of dendrite tip shape selection in a metallic alloy

H. Neumann-Heyme[1a], N. Shevchenko[1a], J. Grenzer[1], K. Eckert[1], C. Beckermann[2b], S. Eckert[1]
[1]*Helmholtz-Zentrum Dresden-Rossendorf (HZDR), 01328 Dresden, Germany*
[2]*University of Iowa, Department of Mechanical Engineering, Iowa City, IA 52242, USA*

The size and shape of the primary dendrite tips determine the principal length scale of the microstructure evolving during solidification of alloys. *In-situ* x-ray measurements of the tip shape in metals have been unsuccessful so far due to insufficient spatial resolution or high image noise. To overcome these limitations, high-resolution synchrotron radiography and advanced image processing techniques are applied to a thin sample of a solidifying Ga-35wt.%In alloy. Quantitative *in-situ* measurements are performed of the growth of dendrite tips during the fast initial transient and the subsequent steady growth period, with tip velocities ranging over almost two orders of magnitude. The value of the dendrite tip shape selection parameter is found to be $\sigma^* = 0.0768$, which suggests an interface energy anisotropy of $\varepsilon_4 = 0.015$ for the present Ga-In alloy. The non-axisymmetric dendrite tip shape amplitude coefficient is measured to be $A_4 \approx 0.004$, which is in excellent agreement with the universal value previously established for dendrites.

## I. INTRODUCTION

Dendritic growth is of vital importance in the formation of microstructures during solidification of metal alloys. The key to understanding dendrite patterns is to study the shape and stability of the dendrite tips, since they stand at the beginning of the morphological evolution [1]. The radius of curvature of a dendrite tip, $R$, represents the initial length scale of a dendrite during growth. The selection criterion [1,2]

$$\sigma^* = \frac{2Dd_0}{R^2 V} \quad (1)$$

provides a fundamental link between the radius $R$ and growth velocity $V$ of a dendrite tip via the dimensionless selection constant $\sigma^*$ that is independent of the growth conditions, where $D$ and $d_0$ are the diffusion coefficient of the melt and chemical capillary length, respectively. According to microscopic solvability theory (MST) [1,2], in a low undercooling regime $\sigma^*$ depends solely on the interface energy anisotropy $\varepsilon_n$ for a $n$-fold crystalline symmetry.

Very near the tip, the shape of a dendrite can be closely approximated as an axisymmetric paraboloid. Further away from the tip, the anisotropic interface energy increasingly affects the shape. Therefore, fins start to develop perpendicular to the growth axis that are located at azimuthal angles where the interface energy has a maximum. Soon after, the growing fins develop transversal instabilities that ultimately lead to the characteristic sidebranches of dendrites. Ben Amar and Brener [3] provided an analytical solution for the universal shape of dendrite tips before sidebranches form. For a fourfold crystalline symmetry, the shape in a plane containing the fins is given by

$$\frac{z}{R} = \frac{1}{2}\left(\frac{x}{R}\right)^2 - A_4 \left(\frac{x}{R}\right)^4, \quad (2)$$

where the coordinates $z$ and $x$ are centered at the tip (cf. Fig. 1c) and have orientations opposite and perpendicular to the growth direction, respectively. The last term in Eq. (2) represents the deviation from the isotropic paraboloidal shape by the fins. The fourth-order amplitude coefficient $A_4$ does not depend on any material properties or growth conditions. Several experimental and numerical studies [4–6] have confirmed that Eq. (2) provides a good approximation to the dendrite tip shape. However, the value of $A_4 \approx 0.004$ found in these studies deviates significantly from the theoretical solution $A_4 = 1/96 \approx 0.01$, the reasons for which are not yet entirely clear [4].

Major progress in understanding the tip shape selection mechanism was achieved during the 1980's and 90's [1]. An important factor in this success was the use of transparent substances in high-precision solidification experiments, e.g. [5–8]. These materials permit *in-situ* observation of dendritic growth features near room temperature using an optical microscope. Nonetheless, the dendrite tip shape selection measurements available for transparent substances do not confirm MST conclusively [1,9,10]. At the same time, the experimental analysis of dendritic microstructure formation in (opaque) metals remained limited to *post-mortem* observations [11]. In a classical analysis of micrographs, Liu et al. [12] determined the selection parameter $\sigma^*$ for a quenched Al-Cu alloy sample. Only with the development of improved synchrotron x-ray facilities and sensors starting in the late 1990's has *in-situ* imaging of dendritic growth in metallic alloys become possible [13]. Although there is a vital interest in extending the experimental validation of dendrite growth theories towards metallic materials, several challenges remain in performing sufficiently accurate measurements in the dendrite tip region. Reliable *in-situ* measurements of dendrite tip shape selection in metals are essentially still missing.

---
[a] Authors contributed equally
[b] Corresponding author: becker@engineering.uiowa.edu; 1-319-335-5681

In transmission radiography there exists a strong trade-off between time resolution and image noise [14]. Even thin layers of metal cause strong absorption of the illuminating x-rays, which inherently limits the number of photons that can be captured by the image sensor. Since the tip region exhibits the strongest dynamics and smallest length scales within a dendritic structure, its imaging is accompanied by significant noise, which prevents a straight-forward use of standard analysis methods. A critical role in the design of x-ray imaging experiments is played by the initial solute concentration of the solidifying alloy. While higher concentrations usually lead to a better image contrast due to larger density differences between the solid and liquid phases, the dendritic structures become finer and more difficult to resolve. When using lower concentrations, the dendrites become larger, but the contrast is reduced. Larger structures are also more prone to confinement effects caused by the limited thickness of the samples.

Despite these difficulties, a limited number of studies have performed *in-situ* radiography measurements of dendrite tips in metals. Several experiments focused on the measurement of tip velocities in Al-Cu alloys, e.g. [15,16], but provided no data on the tip shape. *In-situ* measurements of both the velocity and the radius of dendrite tips have only recently been reported [17,18]. Mirihanage et al. [17] measured time varying tip velocities and radii for directional solidification of an Alb-15wt.%Cu-9wt.%Si alloy and successfully compared measured and predicted solute concentration fields ahead of the dendrite tips. However, dendrite tip shape selection was not examined and no values of $\sigma^*$ and $A_4$ were determined. Clarke et al. [18] measured tip velocities and radii for an Al-Cu alloy and compared their results with phase-field simulations. They found that the measured tip radii essentially follow the $R \sim V^{-1/2}$ relationship from MST [Eq. (1)], but their radii distribution range at a given tip velocity was between a factor of two and three. This large uncertainty was attributed to a lack of contrast in the radiographic images. Consequently, no values of $\sigma^*$ and $A_4$ were obtained. Their phase-field results indicate that in their experiment the effect of melt convection on tip radius selection was negligibly small.

In the present study, *in-situ* measurements of dendrite tip shape selection are made for a hypereutectic Ga-35wt%In metallic alloy, which solidifies near room temperature. It should be noted that the primary phase in this alloy (In) has a body-centered tetragonal (bct) crystal structure, which corresponds to a slightly distorted face-centered cubic (fcc) system with fourfold crystalline symmetry. The evolution of the dendrite tip shape is observed over a large range of growth velocities by high-resolution synchrotron radiography imaging. The region around the tip is extracted by a relatively small window that follows the tip position over time. The image quality is substantially improved using a temporal averaging filter. Another novel aspect of the present study is that the parameter group $Dd_0$ in the definition of the selection parameter, Eq. (1), is obtained directly from previous measurements of the universal pinching dynamics during dendrite sidebranch detachment [19]. This enables the tip selection parameter $\sigma^*$ to be predicted much more accurately, as the measurement of the individual material properties needed to evaluate the group $Dd_0$ usually requires elaborate experimental tests that are available only for a small number of the most frequently analyzed alloys [20–22].

## II. EXPERIMENTAL SETUP AND PROCEDURE

Figure 1a shows the test cell used in the present solidification experiment. The cell was previously employed in a different study carried out by means of a microfocus x-ray tube [23]. The Ga-35wt.%In alloy is prepared from 99.99% Ga and 99.99% In. The alloy is melted and filled into the cell, which has a liquid metal volume of $22 \times 22 \times 0.2$ mm$^3$. As shown in Fig. 1a, the heating/cooling system consists of two sets of Peltier elements in thermal contact with the bottom and top ends of the solidification cell. The Peltier elements are connected to a control unit that allows for the independent adjustment of the cooling rate and temperature gradient across the cell. The distance between the heater and the cooler is approximately 23 mm. Temperatures are measured using two miniature type-K thermocouples that are in thermal contact with the outer surface of the cell near the edge of the Peltier elements. The accuracy of the temperature control is $\pm 0.2$ K. The vertical temperature gradient is calculated from the temperature difference measured between these two thermocouples. The alloy is directionally solidified from top to bottom with a constant temperature gradient of $2 \pm 0.4$ K/mm and cooling rate of 0.002 K/s.

The experiment is performed at the ID19 beamline of the European Synchrotron Radiation Facility (ESRF) in Grenoble, France. The solidifying sample is exposed to a monochromatic, parallel x-ray beam with a photon energy of 40 keV. Conventional transmission radiographs are obtained by means of a scintillator that is coupled to a high speed sCMOS camera (PCO.edge) with $2048 \times 2048$ pixels, yielding an effective pixel size of 0.72 µm. This imaging equipment leads to a field of view of about $1.5 \times 1.5$ mm$^2$ [24]. Radiographs are recorded continuously with a frame rate of $2 \text{ s}^{-1}$. To change the location of the observation window, the position of the solidification cell is manipulated with respect to the x-ray beam by means of a motorized positioning system.

Before solidification is initiated, the Ga-In alloy is heated to a temperature of 70 °C and held at this temperature for a few minutes. The liquidus temperature of the alloy is about 45 °C. During this stage, images are taken by real-time radiography to ensure that the alloy is homogenously mixed before the cooling process is started. Dark field and flat field images of the completely molten alloy are also recorded for subsequent data processing. The cooling of the cell is initiated after recording these reference images. Soon after the first appearance ($t = 0$) of solid alloy at the upper cold end of the sample, the solidification front is searched for dendrites that are suitable for analysis. Dendrites are selected by requiring that the underlying crystallographic orientations are well aligned with the sample plane and the viewing direction, as indicated by the angles of the dendrite sidebranches with respect to the main stem.

The selected dendrite tips are then followed by shifting the cell position in regular intervals along the growth direction to ensure that the tips remain within the field of view.

Figure 1b shows an example stack of images that consists of the last frames captured at seven successive cell positions. Each image is outlined by a different color dashed line. The dashed black rectangle in Fig. 1a. illustrates the approximate position of the image stack of Fig. 1b within the test cell. In the present study, four tips are selected for detailed analysis, as indicated by the labels Tip 1–4 in Fig. 1b. The measured tip trajectories are displayed as solid lines, where the colors correspond to the cell positions. To observe changes in the global dendritic structure at some intermediate times, the tip-following scans are interrupted to perform mesh-scans of the entire solidification cell. This is the reason for the measurement gaps that are apparent in the data presented below.

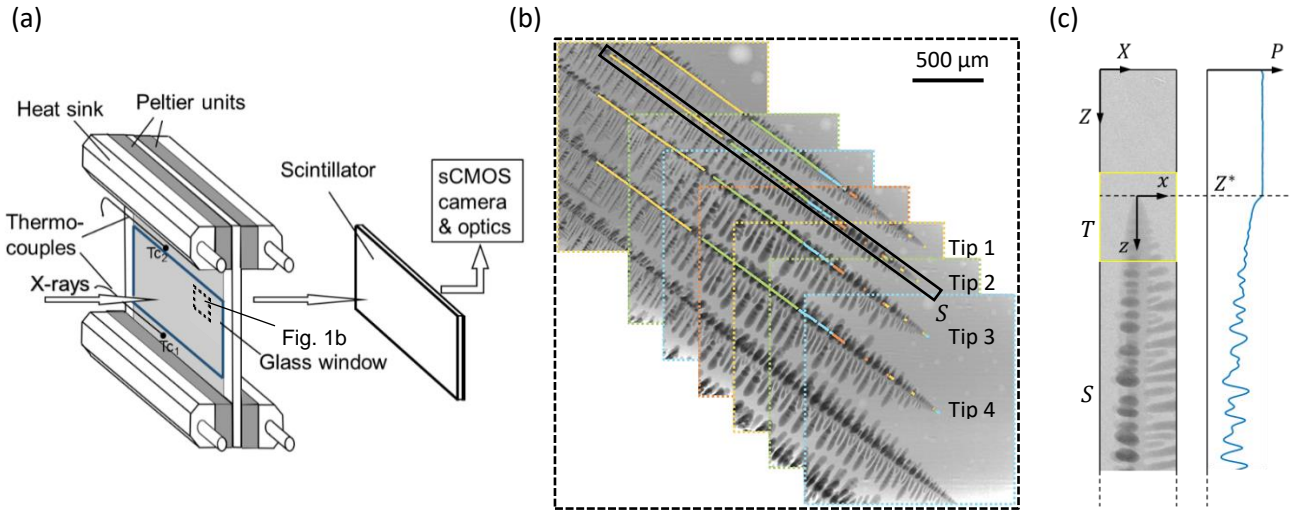

FIG. 1. Experimental setup and image analysis: (a) sketch of the solidification cell [25]; (b) example of a region used in the dendrite tip analysis (dashed rectangle in (a)), where the stack of images represents an overlay of the last frames taken at seven successive cell positions; (c) close-up of the selection frame $S$ and moving frame $T$ for Tip 2 ($t = 748$ s, black rectangle in (b)), minimum projection $P(S)$ and identified tip position $Z^*$.

## III. IMAGE PROCESSING AND ANALYSIS

Noisy images pose a formidable challenge to reliably measuring the fine morphological details that are associated with a dendrite tip. In the present image processing approach use is made of the fact that in a co-moving frame of reference a tip retains a nearly stationary shape for small time intervals up to the distance where sidebranches appear. The principal processing steps described in the following are implemented using the DIPimage toolbox [26] within the MATLAB programming environment.

First, a flat-field correction is performed to eliminate non-uniformities in the image background. A reference background image is obtained as the average over the first 20 frames at a given solidification cell position. Then, for convenience, the background intensity of all frames is normalized to some predefined value $v_B$. This is achieved by scaling of the image contrast in each of the individual frames.

The following processing steps are performed individually for each dendrite tip. A narrow rectangular image region $S(X, Z, t)$ is defined that contains the entire growth path (shown as black outline in Fig. 1b for Tip 2). Figure 1c shows a close-up of the rectangular region near the tip, where its coordinate system $\{X, Z\}$ is fixed to the sample plane. To improve robustness, the image data is slightly smoothed using a spatial Gauss-filter with a standard deviation of one pixel.

Next, the dendrite tip position $Z^*(t)$ is determined as a function of time. The denser solid has a lower intensity than the (liquid) background. Thus, the progression of the tip parallel to the $Z$-coordinate can be tracked by performing a minimum projection $P(Z,t) = \min_{X}[S(X,Z,t)]$ along the lateral direction $X$. In the resulting image $P(Z,t)$, the tip position $Z^*(t)$ is then identified as the border separating the high and low intensity regions. This is illustrated in Fig. 1c, where the profile of $P$ is shown at $t = 748$ s.

The tip velocity is calculated as the time derivative of $Z^*(t)$ using a finite difference approximation followed by locally estimated scatterplot smoothing (LOESS), which is based on second degree polynomials. Here, particular care is taken to properly handle the time gaps in the measurement data that are mentioned in the previous section. Furthermore, $Z^*(t)$ is used to define a small sub-image $T(x,z,t)$ of $S$ that only contains a narrow region around the tip. In this co-moving window, the tip location remains fixed at the origin of the local coordinate system $\{x,z\}$.

Examples of a tip-centered window for Tip 3 at three different times are provided in the top row of Fig. 2a. As shown in Fig. 3a below, these times correspond to a large range of tip growth velocities. As the most important step in reducing image noise, a uniform temporal filter (moving average) is now applied to $T$ over a range of $t \mp 70$ frames ($\mp 35$ s). The resulting images in the center row of Fig. 2a show a very well-defined dendrite tip, indicating that the tip shape is nearly stationary within the chosen time interval for averaging. Since no noise-free reference images are available, the peak-signal-to-noise ratio (PSNR) [26] is estimated for the noisy and time-averaged images. An improvement from about $34.5 \pm 0.4$ dB (top row) to $55.3 \pm 0.1$ dB (center row) is achieved.

The boundary between the dendrite and the background becomes unsharp at some distance below the tip, which is caused by sidebranches that are not stationary in the moving reference frame. The low-noise version of $T$ can be efficiently segmented into dendrite and background regions by applying a constant threshold value slightly below the mean intensity of the background, $v_B$. Both the time range for averaging and the threshold value for segmentation are selected as a compromise between the robustness and accuracy of the final measurement results. The bottom row of Fig. 2a displays the segmentation result as a black contour superimposed on the original image.

Finally, the pixel locations of the tip contour are fitted to the theoretical tip shape given by Eq. (2). Figure 2b illustrates the fitting process for Tip 3 at $t = 2{,}538$ s. The tip shape is fitted in two steps. The tip radius $R$ is measured in Fit 1 using a relatively narrow fit range of 10 pixels (7.2 µm) from the tip, as indicated by the red arrow. In Fit 1 the second term in Eq. (2) is neglected by setting $A_4 = 0$, which corresponds to a purely parabolic shape. Fit 2 is then used to measure the anisotropic deviation from the parabolic shape as quantified by the fourth-order amplitude coefficient $A_4$. For this fit, a larger fit range of $8R$ is applied (blue arrow). This range is still close enough to the tip that the asymmetries in the dendrite shape and sidebranches visible in the lower portion of the images in Fig. 2a do not affect the fit. To suppress unreliable measurements, fitting data that have a coefficient of determination of $r^2 < 0.9$ are neglected. Based on the confidence intervals of the two fits, the mean uncertainties in $R$ and $A_4$ are 5.5% and 33.6%, respectively.

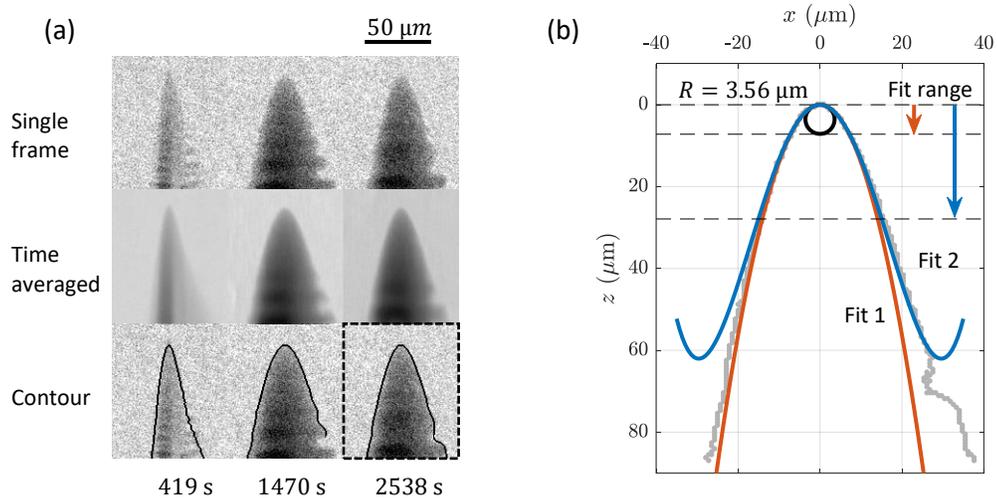

FIG. 2. Dendrite tip image processing and analysis: (a) tip-tracking observation windows showing Tip 3 at three different times (see dashed lines in Fig. 3), where the original images are shown in the top row, the time averaged images in the center row, and the segmentation contours overlaid on the original frames in the bottom row; (b) fitting of the tip shape (gray dots) at $t = 2{,}538$ s (dashed rectangle in (a)), where Fit 1 (red line) is a parabolic fit over a narrow fit range and Fit 2 (blue line) is the anisotropic shape fit using an adaptive fit range of $8R$.

## IV. RESULTS AND DISCUSSION

The four neighboring dendrites selected for analysis (Fig. 1b) are part of a larger grain that had previously nucleated near the cold upper end of the sample. The selected dendrites are tilted at a uniform angle of ∼55° with respect to the vertical sample axis. As can be seen in Fig. 1b, the sidebranches are longer on the downward facing side of the dendrites, where the undercooling is larger. The temperature of the tips decreases from about 37°C to 27°C during the timespan considered in the analysis.

### A. Tip velocity, radius, and shape

The measured growth velocities and radii for the four selected dendrite tips are shown in Fig. 3. The tip velocities vary by almost two orders of magnitude over the roughly 6,000 s long measurement period. Initially, the growth velocity is high because of the relatively large initial undercooling of the melt. The rapid growth results in a relatively small tip radius of about 1 μm. The tips then relax towards a slower, steady growth regime with a larger tip radius. Towards the end of this transition, at about 1,500 s, the tip velocity reaches a local minimum and the tip radius a maximum close to 5 μm. Subsequently, the tip velocity experiences a slight increase and the tip radius a decrease. After about 2,500 s, the tip growth is almost completely steady, except for Tip 1 which continues to slow down.

At the beginning of the measurements, the growth rate is highest for Tip 1 and decreases towards Tip 4. This can be explained by the fact that Tip 1 is closest to the cold top boundary and experiences the largest undercooling. After a short time, however, the velocity ranking among the tips becomes inverted. The trailing tips (lower number) experience a stronger deceleration, and their velocity eventually falls below that of the leading tips (higher number). This may be explained by diffusive interactions between the dendrites. A trailing tip has a reduced undercooling because of solute diffusion from the neighboring dendrite that is ahead. The velocity decrease of Tip 1 after 1,500 s can also be attributed to diffusive interactions.

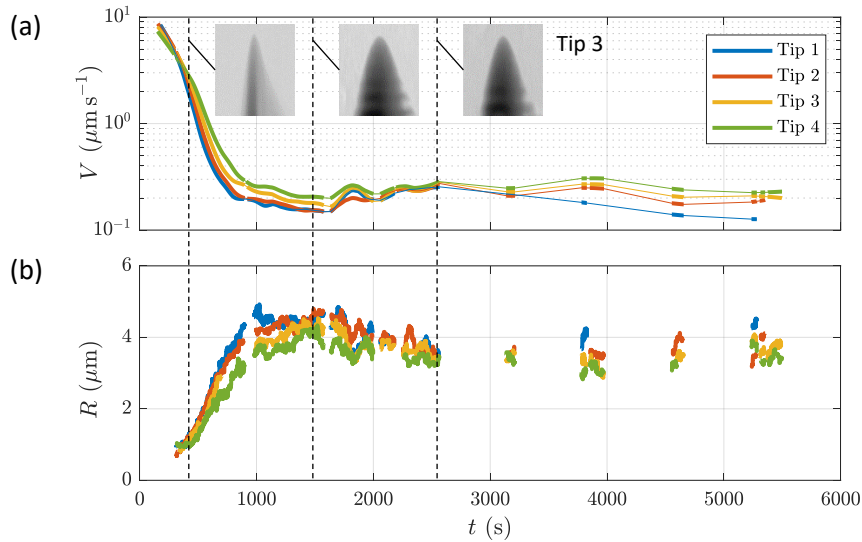

FIG. 3. Measured tip velocities (a) and tip radii (b) as a function of time. Missing tip velocities are bridged by thin lines in (a) to guide the eye.

The measured fourth-order amplitude coefficients $A_4$ are shown in Fig. 4. The results for the four dendrites exhibit no systematic trend with tip number, and differences among them consist of statistical noise only. Therefore, instead of providing the $A_4$ values individually for each of the four dendrites, only a moving average of their median value is shown (solid black line) together with their minimum-maximum range at each measurement time (gray area). As expected from theory, the median $A_4$ values are, despite some fluctuations, essentially constant over time and, therefore, independent of the growth conditions. Density distributions of the $A_4$ values for each tip are provided on the right side of Fig. 4. The four density curves are largely consistent in their shape and the location of their maximum. The density maxima (peaks) are relatively sharp, further confirming that a single value of $A_4$ is preferred throughout the measurements. It can be seen that the density distributions are somewhat skewed towards lower $A_4$ values. These deviations from a symmetrical normal distribution imply that using the mean value and standard deviation of all combined data would likely result in a poor estimate of $A_4$ and its uncertainty. Instead, the fourth-order amplitude coefficient is estimated here based on the location of the peaks in the density distributions. The mean location of the four peaks (dashed line in Fig. 4) is given by $A_4 = 0.00406 \pm 0.00039$, where the uncertainty is the standard error for a 95%-confidence level. Note that this uncertainty is much smaller than the mean uncertainty of a single measurement, which is 33.6% (see above). The present value of $A_4 \approx 0.004$ is in excellent agreement with other measurements and numerical results reported in the literature [4–6]. These previous results are for different alloy systems, indicating that the value of the fourth-order amplitude coefficient is indeed universal. The ability to determine such a small coefficient to reasonable accuracy provides considerable confidence in the present tip shape measurements.

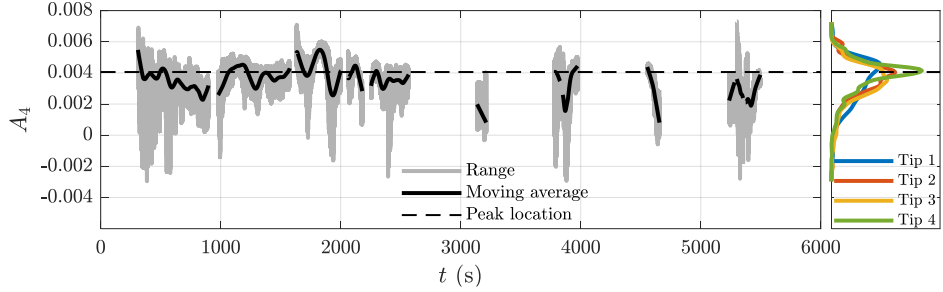

FIG. 4. Measured fourth-order amplitude coefficient as a function of time. The black line is a moving average of the median values, while the gray areas represent the range of the smallest and largest values among the four tips. Density distributions of the $A_4$ values measured for each of the four tips are shown in the plot on the right side (colored lines). The black dashed line represents the mean of the peaks in the density distributions.

## B. Tip selection parameter $\sigma^*$

As indicated by Eq. (1), the determination of the dendrite tip selection parameter $\sigma^*$ requires not only the measurement of the tip velocity and radius, but also the knowledge of the product of the diffusion coefficient and the capillary length, $Dd_0$. The individual measurements necessary to determine this property product are far from trivial and only available for a small number of materials. Recently, some of the present authors proposed a method to obtain $Dd_0$ as a single parameter based on measurements of the pinching dynamics during the detachment of dendrite sidebranches [19]. In this previous study, the product $Dd_0$ was determined for a Ga-25wt.%In alloy at a temperature of approximately 15°C using an experimental setup similar to the present one. A value of $Dd_0 = 0.122 \pm 0.0026\ \mathrm{\mu m^3 s^{-1}}$ was found. To evaluate if the same value can be used for the present experiments involving a Ga-35wt.%In alloy and a temperature range of $27 - 37°\mathrm{C}$, it is necessary to consider the temperature dependence of each property in the product $Dd_0$. Using the definition of the chemical capillary length, $Dd_0$ is given by

$$Dd_0 = \frac{D\Gamma}{|m|(1-k)C_l^{eq}}, \qquad (3)$$

where $\Gamma$ is the Gibbs-Thomson coefficient, $m$ the liquidus slope, $k$ the partition coefficient, and $C_l^{eq}$ the equilibrium solute concentration of the melt at a given temperature. Based on the data summarized in Ref. [19], a careful examination of the temperature dependence of each material property in Eq. (3) revealed that only the diffusion coefficient $D$ and the liquidus slope $m$ show strong variations in the near-eutectic region. However, the combined effect on the product $Dd_0$ is quite small because the individual variations tend to compensate each other in Eq. (3). It is estimated that the maximum deviation from the previously measured value for $Dd_0$ for the current temperature range is less than $\pm 4\%$. Therefore, the value of $Dd_0 = 0.122\ \mathrm{\mu m^3 s^{-1}}$ is adopted in the following. The overall uncertainty in this value is estimated to be 6.1%.

Figure 5 shows the time variation of the tip selection parameter $\sigma^*$ calculated from Eq. (1) based on the measured $V$ and $R$ and the above value for $Dd_0$. The figure is designed the same way as Fig. 4 above. It can be seen that the tip selection parameter is largely constant over time. This result is expected, since theory predicts that $\sigma^*$ is independent of the growth conditions. It is remarkable nonetheless, because the tip growth velocities vary by almost two orders of magnitude over the course of the measurements. The density distributions for $\sigma^*$ are largely consistent among the different tips. Taking the mean location of the peaks of the density distributions for the four tips gives $\sigma^* = 0.0768 \pm 0.0034$, where the uncertainty is again the standard error for a 95%-confidence level. Based on the uncertainties in the tip radius and the product $Dd_0$ provided above, and a negligible uncertainty in the measured tip velocity, the mean uncertainty in a single measurement is estimated to be 12.6%.

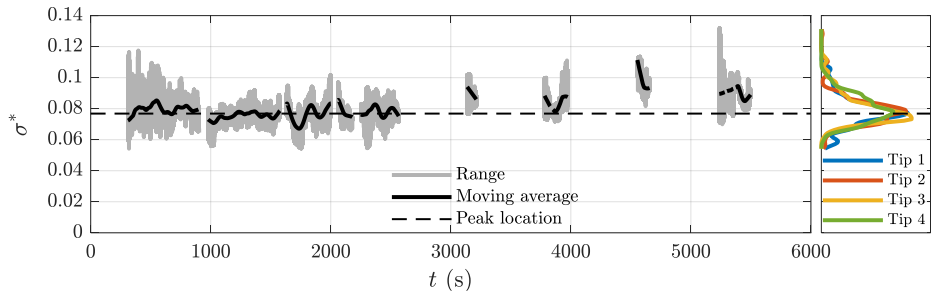

FIG. 5. Measured tip selection parameter as a function of time. The black line is a moving average of the median values, while the gray areas represent the range of the smallest and largest values among the four tips. Density distributions of the $\sigma^*$ values measured for each of the four tips are shown in the plot on the right side (colored lines). The black dashed line represents the mean location of the peaks in the density distributions.

The value for $\sigma^*$ determined here is compared to other experimental results and MST in Fig. 6. The line showing the dependence of the tip selection parameter on the four-fold interface energy anisotropy $\varepsilon_4$ is the linearized MST result in the limit of small tip growth Péclet numbers [27]. The MST curve suggests an interface energy anisotropy of $\varepsilon_4 \approx 0.015$ for the present Ga-In alloy. Unfortunately, no independent measurement of $\varepsilon_4$ for a Ga-In alloy was found in the literature. The $\sigma^*$ measured *post-mortem* for an Al-Cu alloy [12], which is the only other data point available for metals, is in excellent agreement with MST. For illustrative purposes, two representative experimental results are included in Fig. 6 for transparent alloys; a more complete comparison for all available $\sigma^*$ measurements in transparent substances can be found in Refs. [9,10]. The measured $\sigma^*$ for $NH_4Br-H_2O$ [28] falls almost exactly on the MST line. The experimental $\sigma^*$ values reported in Ref. [5] for succinonitrile (SCN) – acetone (ACE) alloys are multiplied by a factor of two to conform to the $\sigma^*$ definition given by Eq. (1). The original $\sigma^*$ values of Ref. [5] correspond to an alternative definition that is uniformly applicable to alloys and pure substances [29]. The $\varepsilon_4$ for SCN was taken from Ref. [22]. The data in Ref. [5] show a strong dependence of $\sigma^*$ on the undercooling, and the vertical bar in Fig. 6 indicates the range of values measured. The relatively large discrepancy between SCN measurements and MST was already noted in Ref. [27], and no explanation has emerged since.

During the present experiment, the diffusion length at the dendrite tips $l_D = D/V$ increases approximately from 50 μm to 1,500 μm. Therefore, the solute diffusion field around the tips will become affected by the sample walls, which are 200 μm apart. Nonetheless, no significant effect on the selection of the tip shape can be noted. This can be explained by the fact that the tip radius, which is the essential length scale in the tip selection problem, remains small compared to the dimensions of the test cell.

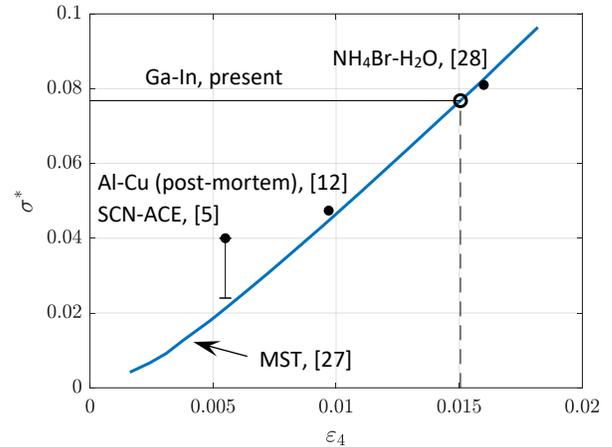

FIG. 6. Dendrite tip selection parameter as a function of the four-fold interface energy anisotropy. The measured $\sigma^*$ (open circle) is superimposed on the line representing MST to indicate the value of $\varepsilon_4$ expected for the present Ga-In alloy.

Moreover, it is likely that melt convection is present in the test cell during solidification. The nature and intensity of such convection would be highly variable since the amount and shape of the solid in the test cell is constantly evolving. Although such convection will cause changes in the dendrite tip growth velocity and radius compared to a purely diffusive environment, it does not affect the dendrite tip shape selection. Otherwise, the tip selection parameter would not be as constant as shown in Fig. 5. This finding agrees with previous theories [30] and experiments in transparent alloys [5] that low to moderate intensity convection has no effect on the selection parameter $\sigma^*$.

IN-SITU MEASUREMENTS OF DENDRITE TIP SHAPE SELECTION IN A METALLIC ALLOY

## V. CONCLUSIONS

*In-situ* measurements of the shape selection of dendrite tips are performed during solidification of the metallic alloy Ga-35wt.%In using a combination of high-resolution synchrotron radiography and advanced image processing techniques. Tip radii down to 1 μm are accurately evaluated. Tip growth velocities range over almost two orders of magnitude. The parameter group $Dd_0$ is obtained from previous measurements of the universal pinching dynamics during dendrite sidebranch detachment. By means of this approach reliable *in-situ* measurements of dendrite tip shape selection in metals could be conducted. A key result of this work is the value of the dendrite tip shape selection parameter, which is found to be $\sigma^* = 0.0768 \pm 0.0034$. Based on microscopic solvability theory, this $\sigma^*$ value suggests an interface energy anisotropy of $\varepsilon_4 = 0.015$ for the present Ga-In alloy. The fourth-order amplitude coefficient describing the non-axisymmetric shape of a dendrite tip is determined to be $A_4 = 0.00406 \pm 0.00039$, which is in excellent agreement with the universal value previously established. The interface energy anisotropy for the present alloy should be measured independently to fully verify MST. The present experiments represent the first reliable direct measurements of dendrite tip shape selection during solidification of metals. Considering the importance of dendritic growth in establishing the microstructure in almost all commercially important metals and alloys [1], and the relatively uncertain nature or limitations of present analytical theories [2,3] and computer simulations [4], the present experiments should provide new motivation for making progress in this field.

## 1. ACKNOWLEDGMENTS

We acknowledge the European Synchrotron Radiation Facility (ESRF) for provision of synchrotron radiation facilities and would like to thank A. Rack and V. Cantelli for assistance in using beamline ID19. We thank O. Keplinger for the continuous support during the experiments as well as R. Weidauer and the HZDR Department of Research Technology for assistance with instruments. This work was financially supported by a recruitment grant of the Helmholtz Association and by NASA (Grant No. 80NSSC20K0828).
[1] W. Kurz, D. J. Fisher, and R. Trivedi, *Progress in Modelling Solidification Microstructures in Metals and Alloys: Dendrites and Cells from 1700 to 2000*, International Materials Reviews **64**, 311 (2019).

[2] D. A. Kessler, J. Koplik, and H. Levine, *Pattern Selection in Fingered Growth Phenomena*, Adv Phys **37**, 255 (1988).

[3] M. Ben Amar and E. Brener, *Theory of Pattern Selection in Three-Dimensional Nonaxisymmetric Dendritic Growth*, Phys. Rev. Lett. **71**, 589 (1993).

[4] A. Karma, Y. H. Lee, and M. Plapp, *Three-Dimensional Dendrite-Tip Morphology at Low Undercooling*, Phys Rev E **61**, 3996 (2000).

[5] A. J. Melendez and C. Beckermann, *Measurements of Dendrite Tip Growth and Sidebranching in Succinonitrile-Acetone Alloys*, J. Cryst. Growth **340**, 175 (2012).

[6] A. Dougherty and M. Lahiri, *Shape of Ammonium Chloride Dendrite Tips at Small Supersaturation*, J Cryst Growth **274**, 233 (2005).

[7] S. Akamatsu and H. Nguyen-Thi, *In Situ Observation of Solidification Patterns in Diffusive Conditions*, Acta Mater **108**, 325 (2016).

[8] W. Huang and L. Wang, *Solidification Researches Using Transparent Model Materials — A Review*, Sci. China Technol. Sci. **55**, 377 (2012).

[9] M. Muschol, D. Liu, and H. Z. Cummins, *Surface-Tension-Anisotropy Measurements of Succinonitrile and Pivalic Acid: Comparison with Microscopic Solvability Theory*, Phys. Rev. A **46**, 1038 (1992).

[10] H. Müller-Krumbhaar, W. Kurz, and E. Brener, *Solidification*, in *Phase Transformations in Materials (Ed. G. Kostorz)* (Wiley-VCH, Weinheim, 2001), pp. 81–170.

[11] M. Gündüz and E. Çadırlı, *Directional Solidification of Aluminium–Copper Alloys*, Mater. Sci. Eng. A **327**, 167 (2002).

[12] S. Liu, J. Li, J. Lee, and R. Trivedi, *Spatio-Temporal Microstructure Evolution in Directional Solidification Processes*, Philos. Mag. **86**, 3717 (2006).

[13] R. H. Mathiesen, L. Arnberg, F. Mo, T. Weitkamp, and A. Snigirev, *Time Resolved X-Ray Imaging of Dendritic Growth in Binary Alloys*, Phys. Rev. Lett. **83**, 5062 (1999).

[14] A. J. Shahani, X. Xiao, E. M. Lauridsen, and P. W. Voorhees, *Characterization of Metals in Four Dimensions*, Mater. Res. Lett. **8**, 462 (2020).

[15] A. Bogno, H. Nguyen-Thi, G. Reinhart, B. Billia, and J. Baruchel, *Growth and Interaction of Dendritic Equiaxed Grains: In Situ Characterization by Synchrotron X-Ray Radiography*, Acta Mater **61**, 1303 (2013).

[16] M. Becker, S. Klein, and F. Kargl, *Free Dendritic Tip Growth Velocities Measured in Al-Ge*, Phys. Rev. Materials **2**, 073405 (2018).

[17] W. U. Mirihanage, K. V. Falch, D. Casari, S. McFadden, D. J. Browne, I. Snigireva, A. Snigirev, Y. J. Li, and R. H. Mathiesen, *Non-Steady 3D Dendrite Tip Growth under Diffusive and Weakly Convective Conditions*, Materialia **5**, 100215 (2019).

[18] A. J. Clarke, D. Tourret, Y. Song, S. D. Imhoff, P. J. Gibbs, J. W. Gibbs, K. Fezzaa, and A. Karma, *Microstructure Selection in Thin-Sample Directional Solidification of an Al-Cu Alloy: In Situ X-Ray Imaging and Phase-Field Simulations*, Acta Mater. **129**, 203 (2017).

[19] H. Neumann-Heyme, N. Shevchenko, Z. Lei, K. Eckert, O. Keplinger, J. Grenzer, C. Beckermann, and S. Eckert, *Coarsening Evolution of Dendritic Sidearms: From Synchrotron Experiments to Quantitative Modeling*, Acta Mater. **146**, 176 (2018).

[20] J. Lee, S. Liu, H. Miyahara, and R. Trivedi, *Diffusion-Coefficient Measurements in Liquid Metallic Alloys*, Metall. Mater. Trans. B **35**, 909 (2004).

[21] P. Savintsev, A. Akhkubekov, K. Getazheev, V. Rogov, and V. Savvin, *Determination of Diffusion and Activity Coefficients in the Gallium-Indium System by the Contact-Melting Method*, Sov. Phys. J. **14**, 467 (1971).

[22] R. E. Napolitano, S. Liu, and R. Trivedi, *Experimental Measurement of Anisotropy in Crystal-Melt Interfacial Energy*, Interface Sci **10**, 217 (2002).

[23] N. Shevchenko, O. Roshchupkina, O. Sokolova, and S. Eckert, *The Effect of Natural and Forced Melt Convection on Dendritic Solidification in Ga-In Alloys*, J. Cryst. Growth **417**, 1 (2015).

[24] P.-A. Douissard et al., *A Versatile Indirect Detector Design for Hard X-Ray Microimaging*, J. Inst. **7**, P09016 (2012).

[25] N. Shevchenko, J. Grenzer, O. Keplinger, A. Rack, and S. Eckert, *Observation of Side Arm Splitting Studied by High Resolution X-Ray Radiography*, IJMR **111**, 11 (2019).

[26] C. L. Hendriks, L. Van Vliet, B. Rieger, G. van Kempen, and M. van Ginkel, *DIPimage: A Scientific Image Processing Toolbox for MATLAB (Ver. 2.9)*, Quant. Imaging Group, Fac. Appl. Sci. Delft Univ. Technol. (1999).

[27] A. Barbieri and J. S. Langer, *Predictions of Dendritic Growth Rates in the Linearized Solvability Theory*, Phys. Rev. A **39**, 5314 (1989).

[28] A. Dougherty and J. P. Gollub, *Steady-State Dendritic Growth of NH4Br from Solution*, Phys. Rev. A **38**, 3043 (1988).

[29] R. Trivedi and W. Kurz, *Dendritic Growth*, Int. Mater. Rev. **39**, 49 (1994).

[30] X. Tong, C. Beckermann, A. Karma, and Q. Li, *Phase-Field Simulations of Dendritic Crystal Growth in a Forced Flow*, Phys. Rev. E **63**, 061601 (2001).